\documentstyle[epsf]{laa}

\begin{document}
\thesaurus{}

   \title{Line Formation in the Atmosphere of Brown Dwarf Gliese~229B~: $CH_4$  
at  2.3 ${\rm \mu m}$}

   \author{Sujan Sengupta  and Vinod Krishan}
   \subtitle{}

   \offprints{Sujan Sengupta (sujan@iiap.ernet.in)}

   \institute{Indian Institute of Astrophysics,  Koramangala,
              Bangalore 560 034, India}

   \date{Received 28 March 2000 ; accepted 12 May 2000}

   \maketitle
 \markboth{Line Formation in Brown Dwarf} {Line Formation in Brown Dwarf }

   \begin{abstract}
We investigate the formation of methane line at 2.3 ${\rm \mu m}$ in Brown Dwarf
Gliese 229B. Two sets of model parameters with (a) $T_{\rm eff}=940 $K and $\log (g)
=5.0$, (b) $T_{\rm eff}=1030 $K and $ \log(g)=5.5$ are adopted both of which provide excellent
fit for the synthetic continuum spectra with the observed flux at a wide range
of wavelengths. In the absence of observational data for individual molecular
lines, we set the additional parameters that are needed in order to model
the individual lines by fitting the calculated flux with the observed flux at
the continuum.  A significant difference in the amount of flux at the core of the
line is found with the two different models although the flux at the continuum
remains the same. Hence, we show that if spectroscopic observation at $2.3{\rm \mu m}$ with
a resolution as high as $R \simeq 200,000$ is possible then a  much better constraint on the surface
gravity and on the metallicity of the object 
could be obtained by fitting the theoretical model of individual
molecular line with the observed data.

   \end{abstract}


\section{Introduction}

The discovery of methane bands in the spectrum of Gl 229B (Nakajima et al 1995,
Oppenheimer, Kulkarni, Matthews \& Nakajima 1995, Geballe, Kulkarni, Woodward, 
\& Sloan 1996) has not only helped to identify  
Gl 229B as a Brown Dwarf but also prompted the creation of the new spectral class of
T dwarfs (Kirkpatrick et al 1999). The synthetic continuum spectra  of the object has been obtained
with and without dust particles (Marley et al. 1996, Griffith, Yelle \& Marley
1998, for a review see Allard et al. 1997). Although the incorporation of condensates can explain the very rapid
decline of the observed continuum flux in the optical region, it is
shown (Tsuji, Ohnaka, \& Aoki 1999, Burrows, Marley, \& Sharp 1999)
 that the pressure broadened red wing of the 0.77 ${\rm \mu m}$ K I doublet
could also account for the observed features of the continuum flux shortward
of 1.1 ${\rm \mu m}$. The spectrum of the first field methane T dwarf
SDSS 1624+0029 shows a broad band absorption feature centered at
7700 $\AA$ which is interpreted (Liebert et al. 2000) as
the K I 7665/7699 resonance doublet. Hence, it is most likely that the shape
of the red spectrum is due to the broad wings of the K I and the Na I doublets
and not due to the presence of condensates.
 All these model spectra together with the bolometric
luminosity of the object (Leggett et al. 1999) and the evolutionary sequences
(Saumon et al. 1996) can constrain the effective temperature of the object very tightly,
however the surface gravity is still poorly constrained. It is shown (Saumon et al. 2000)
that a  multi-parameter fit of the observed spectrum both for the K-band and for the red end is possible with different
metallicities. Therefore, in order to
determine the physical properties of the atmosphere of Gl 229B and the other
T dwarfs  uniquely, more comprehensive theoretical modeling of the observed spectrum
is required. One of the most important studies is the individual line formation
by the most abundant molecules, eg. $CH_4$, $H_2O$, $NH_3$ etc.

  In this paper, we for the first time, attempt to model the line formed
by methane at 2.3 ${\rm \mu m}$ and show that if individual lines of abundant molecules, in
particular that of methane can be resolved observationally, then stringent constraint
 on the surface gravity, on the metallicity and on the temperature at the
bottom of the atmosphere, where the optical depth is very high, can be obtained. 


\section{The Line Transfer Equations}

In order to model the individual line of a particular molecule one needs
to solve the Non-LTE line transfer equations. 
The two level atom line transfer equation in the plane parallel stratified 
medium can be written as (Mihalas 1978)
\begin{eqnarray}
 \pm\mu\frac{\partial  I(\nu,\pm\mu,z)}{\partial z } =  k(\nu,z)
[S(\nu,\pm\mu,z)-I(\nu,\pm\mu,z)]\, , 
\end{eqnarray}
where $\mu$ is the cosine of the angle made by the ray to the normal.
 The total absorption coefficient  
$k(\nu,z)=k_l(z)\phi(\nu,z)+k_c(\nu,z),$ where
$k_l$ is the frequency integrated line absorption coefficient, $\beta(z)=k_c/k_l$, 
$k_c$ is the absorption coefficient for the continuum  and $\phi(\nu,z)$ is
the line profile function. 
The source function is given by
\begin{equation}
S(\nu,\pm\mu,z)=\frac{\phi(\nu,z)S_l(\nu,\pm\mu,z)+\beta(z)S_c(z)}{\phi(\nu,z)
+\beta(z)},
\end{equation}
where $S_c$ is the continuum source function which is set equal to the Planck
function $B(\nu,z)$ and the line source function $S_l$ is written as
\begin{eqnarray}
& & S_l(\nu,\pm\mu,z) = \frac{1-\epsilon}{2}\int_{0}^{\infty}{\phi(\nu',z)d\nu'}\times \nonumber \\
& & \int_{-1}^{1}{R(\nu,\mu:\nu',\mu')I(\nu',\mu',z)d\mu'}+\epsilon B(\nu,z) \, ,
\end{eqnarray}
 where $\epsilon$ is the probability per scattering that  the photon is
destroyed by collisional de-excitation and is given by (Mihalas 1978)
$\epsilon \equiv \epsilon '/(1+\epsilon ');
\epsilon ' \equiv C_{ul}(1-e^{h\nu/kT})/A_{ul},$
$C_{ul}$ is the rate of collisional de-excitation and $A_{ul}$ is the rate of
spontaneous emission.
$R$ is the redistribution function
defined according to a hybrid model prescription by Rees \& Saliba (1982) and
is given by
$R(\nu,\mu;\nu',\mu')=P(\mu,\mu')R(\nu,\nu'),$
where $P(\mu,\mu')$ is the phase function and $R(\nu,\nu')$ is the
angle-averaged frequency redistribution function for isotropic scattering.

\section {The Absorption Coefficient}

Under chemical equilibrium among different atomic and molecular  species
in their standard states, the number density of $CH_4$ for the molecular equilibrium
 reaction $C+2H_2 \rightleftharpoons  CH_4$ can be written as (Sharp 1985)
\begin{equation}
n_{CH_4}=n_Cn^2_{H_2}\left(\frac{h^2}{2\pi m kT}\right)^310^{D_o\Theta}\frac{Q_{CH_4}}{
Q_CQ^2_{H_2}},
\end{equation}
where $Q_C$ and $Q_{H_2}$ are the partition functions for $C$ and $H_2$ , 
 $m$ is the "multiple" reduced mass and
$D_o$ is the dissociation energy of $CH_4$ in eV, $n_C$ and $n_{H_2}$ are the number
densities of $C$ and $H_2$ , $\Theta(z)=5040/T(z)$ where $T(z)$ is in Kelvin, and the other symbols have their
usual meaning. Although chemical equilibrium of $C$ and $H_2$ should be coupled
with some other molecules, dominance of $CH_4$ molecule in the atmosphere
would make the above equilibrium the most probable one. Hence, we ignore,
for the present, chemical equilibrium of $C$ and $H_2$ with other molecules. 
For the purpose of calculating the number densities of the molecules in various
energy levels, we assume that the lower atmosphere of the object is in local
thermodynamic equilibrium (LTE). Then the number of particles $n_J$, in a specified
rotation level $J$ is related to the total number of particles in all levels, $n_{CH_4}$,
according to the following relation (Larson 1994):
\begin{equation}
\frac{n_J}{n_{CH_4}}=\frac{(2J+1)e^{-E_J/kT(z)}}{Q_{CH_4}},
\end{equation}
where (2J+1) is the degeneracy factor and $e^{-E_J/kT}$ is the Boltzmann factor. 
Now the integrated line absorption coefficient can be written as
\begin{eqnarray}\label{lineabs}
k_l(z) & = & \frac{\pi e^2}{m_e c}fn_J(z)=
 \frac{\pi e^2}{m_e c}f\left[\frac{h^2}{2\pi m k T(z)}\right]^{3}10^{D_o\Theta(z)}\times \nonumber \\
& &\frac{(2J+1)e^{-E_J/kT(z)}}{Q_CQ^2_{H_2}}n_C(z)n^2_{H_2}(z)\, ,
\end{eqnarray}
where $f$ is the oscillator strength of the transition.
We calculate the Q's for $H_2$ and $C$ using polynomial expressions given by
Sauval \& Tatum(1984). For a small value of $J$, the Boltzmann factor
reduces to 1. The dissociation energy $D_o$ of $CH_4$ is taken to be 4.38 eV
(Jorgenson 1994).
We use the continuum absorption coefficient $k_c$ provided by D. Saumon.

\section{The Synthetic Continuum Spectra in the Infra-red Region}

We present the synthetic continuum spectra of Gl 229B (Figure~1) in the infra-red region
where the signature of methane is clear and dominant. The monochromatic
radiative transfer equations are solved numerically by using discrete
space theory (Peraiah and Grant 1973). We have adopted two
sets of model parameters, model (a): T$_{\rm eff}$=940 K, log g=5.0 (g in cms$^{-2}$) and
[M/H]=$-0.3$ (K band) and model (b): $T_{\rm eff}$=1030 K, log g=5.5  and
[M/H]=$-0.1$ (K band). The temperature and pressure profiles for both the models
are obtained from M. Marley (private communication). The values of $T_{\rm eff}$ and the surface gravity g are
constrained by the bolometric luminosity of Gl 229B and the evolutionary
sequence of Saumon et al (1996). These synthetic spectra  fit 
the entire observed spectrum of Gl 229B except in the near infra-red
region. It should be mentioned that a good fit with the observational data
for the entire wavelength region
is not possible with the same value of the metallicity and the surface gravity. 
The above sets of model parameters are two
of many optimal sets of parameters that can produce good fit.  In order to match the observed flux 
shortward of 1.1 ${\rm \mu m}$ that declines very rapidly, either dust particulates or
alkali metals have to be incorporated. However, the size of the grain that is needed to
explain the observed spectrum in the optical region is too small to play
any role in the infra-red region we are interested in the present work. Hence 
in this spectral region
the law of Mie scattering that describes the angular distribution of
photons in a dusty atmosphere reduces to that of Rayleigh scattering. Therefore, 
the synthetic spectrum in this region matches well with the observed flux
when Rayleigh's law of scattering is used.
Recent ovservational evidence (Liebert et al. 2000) implies that there is
no compelling reason to introduce dust or additional opacity source in the
atmosphere of methane dwarf. Therefore in the present work we have not
incorporated dust opacity. 

  Figure~1 shows that there is no difference in the calculated continuum flux
with the two sets of model parameters. Both the models fit the observed data
very well. 
   \begin{figure}[htbp]
{\epsfxsize=9cm \epsfysize=6cm \epsfbox{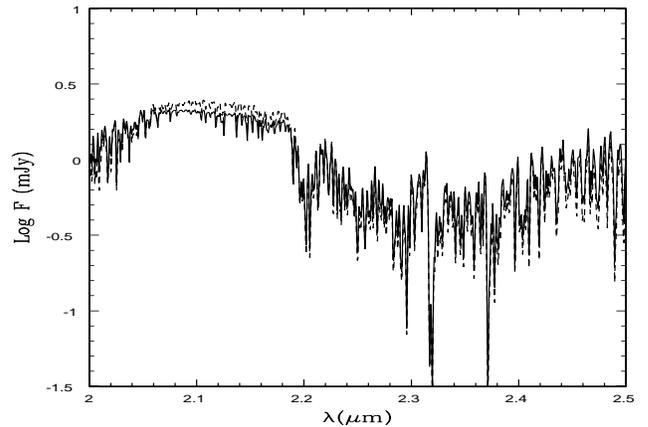}}
\caption{Synthetic continuum spectrum of Gl 229B:  broken line is for the model
(a) with $T_{\rm eff}=940 $K and $log(g)=5.0$ (g in ${\rm cms^{-2}}$), solid line for the model (b) with
$T_{\rm eff}=1030 $K and $log(g)=5.5$}
\label{fig1}
 \end{figure}

\section{Results and Discussion}

The continuum opacity and the temperature-pressure profile of the atmosphere are set by matching
the synthetic continuum spectrum with the observed spectrum for the entire
wavelength region. The modeling of individual lines needs the values of additional
parameters. In the absence of observational data for individual lines, we 
set the value of these parameters by matching the calculated flux in the
continuum with the observed continuum flux at 2.3 ${\rm \mu m}$. Since the line is
very weak in intensity, we choose the profile function as (Mihalas 1978)
\begin{equation}
\phi(\nu,z)=\frac{1}{\sqrt{\pi}\Delta\nu_D(z)}e^{-[(\nu-\nu_0)/\Delta\nu_D(z)]^2},
\end{equation}
where $\Delta \nu_D$ is the thermal Doppler width and $\nu_0$ is the line
center. It should be worth mentioning that individual molecular lines are
usually not saturated enough so that pressure broadening is less important
for them as compared to strong atomic lines. Moreover, molecular lines often overlap so strongly that
their wings are completely masked (Schweitzer et al. 1996) and only the
Gaussian line cores of the strongest molecular transitions are observed.
The atmosphere of a brown dwarf is therefore only weakly sensitive to the
Van der Walls damping constant. 
Nothing is known, at present, about the rotation of the brown dwarf Gl 229B 
around its own axis of rotation. 
If the projected velocity `$v\sin i$' of the object is greater than
2 to 5 ${\rm km s^{-1}}$ then rotational broadening could be
significant. However, in the present work we have neglected rotational
broadening in order to make the results consistent with the 
calculation of the  evolutionary sequences by Saumon et al. (1996) that
constrains the surface gravity and the effective temperature of the object.
The whole purpose of the present work is to show that with different values
of the surface gravity and the metallicity, the flux at the line core is significantly different
although it is the same in the continuum. Since rotational broadening would
affect the spectrum equally for both the models, it is not important in the context
of the present work. We assume complete frequency redistribution and use
Rayleigh phase function for the angular redistribution. 

  The parameters that are to be set in order to model the $CH_4$ line at 2.3 
${\rm \mu m}$ are $\epsilon$, $f$, $n_C(z)$, $n_{H_2}(z)$, and the degeneracy factor
$(2J+1)$. We define $s=(2J+1)fn_C(z)n^2_{H_2}(z)$ guided by equation (\ref{lineabs})
and set the values of $s$ and $\epsilon$ such that the calculated value  of the
flux at the continuum matches with the observed continuum flux. We assume 
that $\epsilon$ is independent of the geometrical depth. This is valid if
the $CH_4$ line formation is confined to a narrow region in the atmosphere. After testing
several empirical laws 
we adopt the usual inverse square law with respect to the geometrical depth
for the variation of the number density of $C$ and $H_2$.
This should be verified when 
observational data becomes available. Experimental determination of the oscillator
strengths for different transitions at 2.3 ${\rm \mu m}$ and at the relevant
temperature could further constrain the number density of $C$ and $H_2$ 
and hence the abundance of methane in the atmosphere of Gl 229B provided
chemical equilibrium exists between the molecule and the atoms. We have solved
the radiative line transfer equations by using discrete space theory (Peraiah 1980). The
theoretical models for the methane line at 2.3 ${\rm \mu m}$ are presented in
Figure~2 . For the model (a) we find that the
flux at the continuum matches with its observed value when $\epsilon=0.08$ and 
$s=(2J+1)fn_1n^2_2=5.9\times 10^{37}$ where $n_1$ and $n_2$ are the number
densities of $C$ and $H_2$ respectively at the bottom of the atmosphere. The
moderately high 
value of $\epsilon$ is consistent with the temperature at the lower atmosphere
where the lines are formed. For the model (b)
the flux at the continuum matches with the observed flux when $\epsilon=0.08$
and $s=2.7\times 10^{39}$. Since the temperature of the lower atmosphere for
the two models does not differ much, the value of $\epsilon$ remains the same for
both the models. However $s$ differs substantially for the two models.
This is because of the fact that for the model (a) the
continuum opacity is less and therefore one has to reduce the line opacity
in order to keep the right ratio ($\beta$) between them so that the calculated flux
matches with the observed flux at the continuum. However, the higher value of
$s$ in the model (b) makes the line opacity higher than that
for the model (a) and so substantial decrease in the calculated
flux at the line core is obtained. The difference in the flux reduces as we go towards
the wings. It is
worth mentioning that the wings could be masked by other lines whereas the
continuum is overlapped by several lines.
   \begin{figure}[htbp]
{\epsfxsize=9cm \epsfysize=6cm \epsfbox{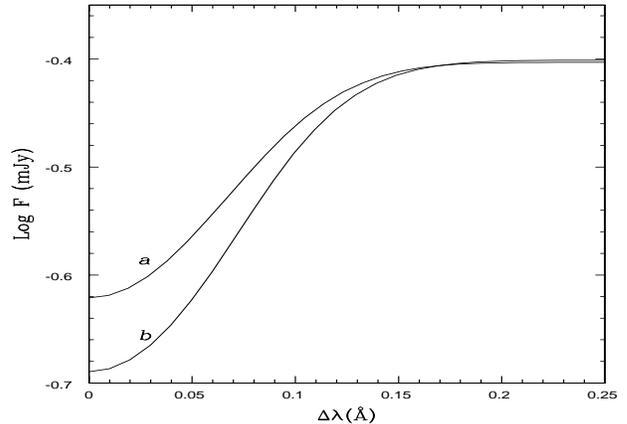}}
\caption{Emergent flux against wavelength from the line center ($2.3 {\rm\mu m}$):
 the curve `a' is for the model (a) and the curve `b' is for the model (b)}
\label{fig2}
 \end{figure}

Figure~2 shows that a spectral resolution as high as 200,000 at $2.3 {\rm \mu m}$
is needed in order to investigate the individual molecular lines. This may
be possible with an appropriate combination of the telescope and the
instrument. For example, the Cooled Grating Spectrometer 4 (CGS4) available
in UKIRT (United Kingdome Infra-Red Telescope)
has a spectral resolution upto 40,000. If observation of Gl 229B is
possible at present  by UKIRT with the maximum resolution power of CGS4 then keeping
the signal to noise ratio (which is proportional to the diameter of the telescope
and to the square root of the integration time of exposure) unaltered, a 10 m telescope (such as Keck I)
can obtain the desired resolution by using a similar type of spectrometer
provided the resolution of the instrument is increased by about five times and the integration
time of exposure is increased by about 2.5 times.

 It is found that the numerical values of $\epsilon$ and $s$ are
very much sensitive to the emergent flux.
The difference in the flux at the line core is clearly due to the different
values of the surface gravity and the metallicity. Theoretical modeling of
 the continuum flux provides a few possible combinations of the metallicity,
surface gravity and the effective temperature that are appropriate in explaining 
the observed continuum flux. The observational fit of the flux at the line core
would decide which one of these combinations should describe the physical
properties of the atmosphere. The physical parameters that are needed to model
the individual molecular line will also be fixed once observational data
is available. 
 Therefore, in conclusion we would
like to emphasize that a theoretical fitting of the observed flux for the 
individual lines of any of the dominant molecules, especially methane, would
not only provide a much better understanding on the abundance of that
molecule and the temperature of the lower atmosphere but also improve
the constraints on the value of the surface gravity of brown dwarfs.

\begin{acknowledgements}
We thank M. Marley for kindly providing the temperature-pressure profiles and
D. Saumon for the continuum opacity data. We also thank the referee for
valuable suggestions and constructive comments. 
Thanks are due to N. K. Rao, T. P. Prabhu, T. N. Rengarajan, K. E. Rangarajan
and G. Pandey for useful discussions.
\end{acknowledgements}

\end{document}